\documentclass[twocolumn,letterpaper]{article}
\usepackage{amsmath, a4wide, amssymb, mathrsfs, mcite, bbm, abstract}
\usepackage[english]{babel}
\usepackage[totalwidth=530pt, totalheight=735pt]{geometry}

\newcommand{\vect}[2]{\ensuremath{{\mathbf #1}_{#2}}}
\newcommand{\vectr}[2]{\ensuremath{{\boldsymbol #1}_{#2}}}
\title{{\bfseries \large Non-Viability of a Counter-Argument to Bell's Theorem}}
\author{\normalsize Marc Holman\thanks{{\itshape \normalsize E-mail :} {\ttfamily \normalsize m.holman@phys.uu.nl}}\\
{\itshape \normalsize Department of Physics and Astronomy, Utrecht University, Princetonplein 5, 3584 CC Utrecht, The Netherlands}}
\date{}

\begin{document}
\twocolumn[
\maketitle
\renewcommand{\abstractname}{}
\begin{onecolabstract}
\noindent It is demonstrated that a recently suggested model for the EPR-Bohm spin 
experiment, based on Clifford algebra valued local variables and observables, 
runs into very serious difficulties and can therefore not be taken as constituting 
a viable counter-example to Bell's theorem.
\end{onecolabstract}
]\saythanks
\noindent In a recent paper \cite{Christian}, Christian has proposed a local realistic model 
for the EPR-Bohm spin experiment, in which the hidden variable space is taken to be a 
subset of the Clifford algebra, $Cl(3)$, of Euclidean 3-space,
$\mathbb{E}^3$, rather than $\mathbb{E}^3$ itself, as in Bell's local realistic model \cite{Bell}.
More specifically, the hidden variable is taken to be the unit trivector,
$\vectr{\mu}{} = \pm I$, where $I$ denotes the (right-handed) unit pseudoscalar, i.e. $I= \vect{e}{x} \vect{e}{y} \vect{e}{z}$, 
with $\{ \vect{e}{x}, \vect{e}{y}, \vect{e}{z} \}$ a right-handed orthonormal basis for $\mathbb{E}^3$
and $I^2 = -1$. Here, the notation of Ref. \cite{Christian} is adopted (note
that although a trivector, \vectr{\mu}{} actually is a ``scalarlike'' quantity,
as it commutes with every element of $Cl(3)$). In physical terms, the hidden
variable of each spin-$1/2$ particle thus represents the orientation of an
orthonormal triad and effectively is just an element of $\mathbb{Z}_2 \simeq \{ \pm 1 \}$.
Now, despite the fact that the model \emph{formally} manages to reproduce 
some quantum theoretical expectation values correctly (see also below), this effective 
$\mathbb{Z}_2$-valued character of the hidden variable is likely to lead to some 
suspicions about the model's prospects to yield the same results as ordinary 
quantum theory in all relevant cases. These suspicions are confirmed when comparing the predictions
of the model for the outcomes of \emph{individual} spin experiments with those of ordinary
quantum theory. The observable, $A_{\vect{n}{}}(\vectr{\mu}{})$, that represents 
the measurement outcome of a spin-meter (i.e. Stern-Gerlach apparatus) $A$ in 
the direction of the unit vector \vect{n}{}, with the particle that entered that 
meter in the hidden variable state \vectr{\mu}{}, is defined within the model as 
(cf. Eq. (16) of Ref. \cite{Christian})
\begin{equation}\label{defobs1}
A_{\vect{n}{}}(\vectr{\mu}{}) \; := \; \vectr{\mu}{} \cdot \vect{n}{} 
\end{equation}
Although the right-hand side of Eq. (\ref{defobs1}) has the appearance of 
a scalar quantity, $\vectr{\mu}{} \cdot \vect{n}{}$ actually is a bivector - 
alternatively, modulo a sign, it is the (Hodge) dual of an ordinary vector, 
i.e. \vect{n}{}. But from a strictly experimental perspective, all that is required
of the spin-meter observable is that it be a quantity that represents measurement
outcomes in terms of $+$ and $-$ \emph{relative} to the direction \vect{n}{}
in which spin is measured and for this purpose the definition $A'_{\vect{n}{}}(\vectr{\mu}{}) := \vectr{\mu}{} I^{-1} \, \mbox{sgn}(n_{\ast})$, 
with sgn and $n_{\ast}$ denoting respectively the sign function and the first 
nonzero component of \vect{n}{} with respect to the basis $\{ \vect{e}{x} , \vect{e}{y} , \vect{e}{z} \}$,
already suffices, as this definition leads to the same outcomes for individual measurements as the definition
(\ref{defobs1}). The sole purpose of the extra degree of freedom introduced in
the definition (\ref{defobs1}) is to serve as a vehicle to circumvent Bell's
theorem. As will be seen shortly however, it actually fails at this since
the definition (\ref{defobs1}), when extended to more than one particle, either
does not correctly describe the correlations between two particles in an EPR-Bohm 
context or leads to other types of inconsistency with experiment.
But there is already a serious difficulty with the definition of the spin-meter
observable to correctly represent the outcomes of repeated measurements on a
single particle in orthogonal directions.
To see this, suppose for definiteness that a spin-measurement is carried out in the
\vect{e}{z}-direction and that the particle which enters the meter is in the hidden
variable state $\vectr{\mu}{}= +I$, so that $A_{\vect{e}{z}}(I) = I \vect{e}{z}$,
i.e. the measurement outcome is ``spin up'' in the positive $z$-direction (it
would of course also be possible to have $I \vect{e}{z}$ correspond to the
``down'' leg of a Stern-Gerlach apparatus oriented in the positive $z$-direction, 
so the present choice - and mutatis mutandis for the other directions - is merely
the most natural one; in any case, this choice does not affect the conclusions
to be derived below). A second spin measurement is subsequently performed on the particle, but the
experimenter now has the choice to measure spin in \emph{either} the \vect{e}{x}-direction
\emph{or} the \vect{e}{z}-direction. Because \vectr{\mu}{} is a \emph{local deterministic}
hidden variable, its value cannot depend on the choice made by the experimenter.
If this value is left unchanged by the first measurement, performing the second
measurement in the \vect{e}{x}-direction would result in ``spin up'' in this 
direction with certainty, in contradiction with the usual quantum predictions. 
But if the value of \vectr{\mu}{} changes as a result of the first measurement,
such that it would yield the outcomes ``spin up'' and ``spin down'' in the
\vect{e}{x}-direction with equal probabilities, this would lead to the wrong
statistics if the second measurement were in fact carried out in the \vect{e}{z}-direction. 
Note that the spin-meter observable in Bell's model \cite{Bell},
as defined in Eq. (9) of that reference, suffers from the same difficulty (as
may e.g. be seen by subsequently performing measurements in the positive $z$-
and $x$-directions, selecting the outcomes $++$ and then performing a measurement
in the positive $z$-direction again), but that the alternative definition in Eq. (4) does not - alternatively, with the definition
(9) one could presumably circumvent the difficulty by postulating that a spin 
measurement in the \vect{n}{}-direction changes the value of \vectr{\lambda}{},
such that it is described by a uniform probability distribution over the upper 
hemisphere with \vect{n}{} as the North Pole.
An objection sometimes raised against the above types of thought experiment, is
that they contain an assumption of ``free choice'' on the part of the experimenter.
As has been noted elsewhere however, if this assumption is in fact not made,
it would be hard to take science seriously, as \emph{Nature could be in an
insidious conspiracy to ``confirm'' laws by denying us the freedom to make
the tests that would refute them} \cite{ConKoch}. Stated somewhat differently,
without the assumption it would be necessary to radically distinguish the concept 
of man-independent, ``ontological reality'' from that of man-dependent, ``empirical
reality'', although it is somewhat ironic that some authors have argued for the
absolute necessity of a distinction between these concepts, precisely because of the nonlocality
that must be part - as a result of Bell's theorem and the experimental results - 
of any sensible notion that may be formed of ``ontological reality'' and the absence 
of such nonlocality in \emph{most} of the phenomena composing ``empirical reality'' 
\cite{dEspagnat}. It seems however that if a theory could be shown to exist that 
could provide a convincing, presumably testable explanation of how an objective, effectively local 
``classical'' world could emerge from a more fundamental one that is in some
important ways nonlocal, it would not be imperative to distinguish between the
two notions of reality. In any case, without the free will
assumption, there is no need to resort to Clifford algebra valued observables
to construct local realist models for quantum theory, and it does at any rate
not appear that acknowledging the existence of two distinct conceptions of reality 
is part of the official programmes of those seeking to construct such models.\\
Apart from the difficulty associated with spin measurements on a single particle,
a second major difficulty with the model is that the definition 
of the observable $B_{\vect{n}{}}(\vectr{\mu}{})$, corresponding to a second, distant spin-meter 
$B$ - also given in Eq. (16) of Ref. \cite{Christian} - does not take into account the experimental 
fact that the observables $A_{\vect{n}{}}(\vectr{\mu}{})$, $B_{\vect{n}{}}(\vectr{\mu}{})$ 
should have opposite signs for the outcome of a single EPR-Bohm spin experiment with both meters set in the
same direction \vect{n}{}. Consistency with experiment thus demands that (see 
also Eq. (13) in Ref. \cite{Bell})
\begin{equation}\label{defobs2}
B_{\vect{n}{}}(\vectr{\mu}{}) \; = \; - A_{\vect{n}{}}(\vectr{\mu}{})
\end{equation}
The product $A_{\vect{a}{}}(\vectr{\mu}{}) B_{\vect{b}{}}(\vectr{\mu}{})$ (cf.
Eq. (17) in Ref. \cite{Christian}) now picks up an extra sign, resulting in the wrong
sign for its expectation value in Eq. (19). 
The only way to avoid this conclusion would be to artificially \emph{interpret}
the measurement outcomes of the $B$-meter in precisely the opposite way; i.e.
to have $B_{\vect{n}{}}(I) = I \vect{n}{}$ correspond to the ``down'' leg of
a Stern-Gerlach apparatus $B$ oriented in the direction \vect{n}{}, if
$A_{\vect{n}{}}(I) = I \vect{n}{}$ corresponds to the ``up'' leg of a Stern-Gerlach
apparatus $A$ oriented in the same direction, and similarly for $\vectr{\mu}{} = - I$. 
Although unnatural, such an alternative interpretation would indeed lead to the correct expectation
values in the current context. However, it is not possible to consistently extend this
way of reasoning to describe correlations between more than two particles.
For instance, in order to yield the correct expectation value for
$A_{\vect{e}{z}}(\vectr{\mu}{}) C_{\vect{e}{z}}(\vectr{\mu}{})$ when a system
of three spin-$1/2$ particles is prepared in the product state $| \, + \, - \, + \rangle$, 
with $+$ and $-$ respectively referring to ``up'' and ``down'' in the \vect{e}{z}-direction,
the observable $C_{\vect{n}{}}(\vectr{\mu}{})$ of the third particle would have 
to satisfy $C_{\vect{e}{z}}(\pm I) = \mp I \vect{e}{z}$, with $- I \vect{e}{z}$ 
corresponding to the ``up'' leg of the apparatus. This would however lead to an 
incorrect expectation value for $B_{\vect{e}{z}}(\vectr{\mu}{})C_{\vect{e}{z}}(\vectr{\mu}{})$.
Thus, the model is unable to take into account the empirical constraints imposed 
by spin systems involving more than two particles, despite the somewhat
intriguing fact that it \emph{formally} manages to exactly reproduce the correct 
quantum expectation values in the case of two particles (note however that the
ordinary quantum observables in the present context are also effectively Clifford
algebra valued, since the algebra generated by $\mathbbm{1}$ and $- i$ times the Pauli matrices
is just the quaternion algebra, $\mathbb{H}$, and $Cl(3) \simeq \mathbb{H} \oplus \mathbb{H}$).
But is it possible to prove that \emph{any} local
realistic model based on non-commuting spin-meter observables is non-viable ?
In order to be consistent with the usual quantum predictions any such observables, 
$A(\vect{a}{}, \lambda), B(\vect{b}{}, \lambda)$, with $\lambda$ denoting a generic 
hidden variable, should satisfy $B(\vect{a}{}, \lambda) = - A(\vect{a}{}, \lambda)$ 
(again taking the EPR-Bohm case, for definiteness, and keeping in mind the previous
discussion) and
\begin{equation}
\int \, d \lambda \rho (\lambda) \, [ A(\vect{a}{}, \lambda) \, , \, B(\vect{b}{}, \lambda) ] \; = \; 0 \; \: , \; \: \int \, d \lambda \rho (\lambda) A^2(\vect{a}{}, \lambda) \; = \; 1
\end{equation}
for all \vect{a}{} and \vect{b}{}. It is not immediately obvious that one could 
prove such models to be inconsistent or to contradict quantum theory as a matter
of general principle. But what if no such ``in principle'' proof is found ?
Or what if a proof is found that rules out only the associative algebras of
observables (the Clifford product and its derived products are associative,
but even here some care is required as the cross product $\times$ between two 
ordinary vectors, related to the derived Clifford wedge product $\wedge$ by duality, is \emph{not} 
in general associative) ? Should that then be taken as constituting a 
\emph{genuinely plausible} loophole for Bell's theorem ? As already noted, from a strictly experimental point of view
all that is required of a spin-meter observable is that it represents measurement
outcomes in terms of $+$ and $-$ relative to the direction in which spin is measured
and for this purpose it is sufficient that it is a real-valued scalar quantity.
The choice to take instead a non-scalar quantity introduces an extra degree of
freedom, something for which there is at present no basis in physical theory.
It is not uncommon to encounter the view that some radical change is required
in conventional physical theory in order to resolve the problem of unifying
quantum theory and general relativity. Yet, even though it is of course impossible
to predict with absolute certainty what the nature of this change will ultimately
turn out to be - assuming a radical change is indeed needed - it is highly questionable 
that it will involve a simple return to the local worldview of classical physics.

\vspace{2ex}

\noindent I thank Dennis Dieks for some very helpful remarks.

\vspace{-2ex}
\bibliographystyle{mltplcit}
\bibliography{bellv2}

\end{document}